\documentclass[prd,superscriptaddress,reprint,aps,amsmath,amssymb,tightenlines,showpacs,nofootinbib]{revtex4-1}

\usepackage{graphicx}
\usepackage{dcolumn}
\usepackage{mathtools}
\usepackage{bm}
\usepackage[dvipsnames]{xcolor}
\usepackage{ulem}
\usepackage[colorlinks=true,citecolor=blue,linkcolor=blue]{hyperref}
\usepackage{comment}

\newcommand\myshade{85}
\colorlet{mylinkcolor}{PineGreen}
\colorlet{mycitecolor}{BrickRed}
\colorlet{myurlcolor}{violet}
\hypersetup{
  linkcolor  = mylinkcolor!\myshade!black,
  citecolor  = mycitecolor!\myshade!black,
  urlcolor   = myurlcolor!\myshade!black,
  colorlinks = true
}

\def\beq{\begin{equation}\begin{aligned}}
\def\eeq{\end{aligned}\end{equation}}

\newcommand{\be}{\begin{equation}} 
\newcommand{\ee}{\end{equation}}

\newcommand{\bk}{\mathbf{k}}

\newcommand{\bs}[1]{\mathbf{#1}}

\begin{document}

\title{General Constraints on Isocurvature from the CMB and Ly-$\alpha$ Forest} 

\author{Matthew~R.~Buckley}
\email{mbuckley@physics.rutgers.edu}
\affiliation{NHETC, Department of Physics and Astronomy, Rutgers University, Piscataway, NJ 08854, USA}
\author{Peizhi~Du}
\email{dupeizhi@ustc.edu.cn}
\affiliation{CAS Key Laboratory of Microscale Magnetic Resonance and School of Physical Sciences, University of Science and Technology of China, Hefei 230026, China}
\affiliation{Anhui Province Key Laboratory of Scientific Instrument Development and Application, University of Science and Technology of China, Hefei 230026, China}
\author{Nicolas~Fernandez}
\email{nico.fer@rutgers.edu}
\affiliation{NHETC, Department of Physics and Astronomy, Rutgers University, Piscataway, NJ 08854, USA}
\author{Mitchell~J.~Weikert}
\email{mjw283@physics.rutgers.edu}
\affiliation{NHETC, Department of Physics and Astronomy, Rutgers University, Piscataway, NJ 08854, USA}

\begin{abstract}
Current cosmological data are well-described by the Lambda-Cold Dark Matter ($\Lambda$CDM) model, which assumes adiabatic initial conditions for the primordial density perturbations. This agreement between data and theory enables strong constraints on new physics that generates isocurvature perturbations. Existing constraints 
typically assume a simple power law form for the isocurvature power spectrum. However, many new physics scenarios -- such as cosmological phase transitions and gravitational particle production -- can deviate from this assumption. To derive general constraints which apply to a wide variety of new physics scenarios, we consider four types of isocurvature modes (dark matter, baryon, dark radiation and neutrino density isocurvature) and parametrize the isocurvature power spectrum using two general forms: a delta function and a broken power law. Using data from the cosmic microwave background (CMB), baryon acoustic oscillations, the Lyman-$\alpha$ forest, and CMB spectral distortions, we place constraints on the isocurvature power spectrum across a wide range of scales, from $10^{-4}\,\textrm{Mpc}^{-1}$ to $10^{4}\,\textrm{Mpc}^{-1}$.
\end{abstract}

\maketitle

\section{Introduction}
\label{sec:intro}

At present, cosmological data from a wide variety of sources, from the Cosmic Microwave Background (CMB) to large-scale structure and galaxy formation, can be well-described by the standard Lambda-Cold Dark Matter ($\Lambda$CDM) model. Within this model, the initial conditions for the primordial density fluctuations are adiabatic~\cite{Planck:2018vyg}. However, some extensions of this model add new physics in the early Universe with isocurvature initial conditions. Such models typically have some component of the initial perturbations which are not sourced by the inflaton.

Current constraints on isocurvature~\cite{Planck:2018jri} typically assume its power spectrum follows a simple power law as a function of perturbation wavenumber, as is naturally expected in axion or curvaton models (for reviews, see e.g., \cite{Marsh:2015xka,Mazumdar:2010sa}). However, numerous well-motivated new physics scenarios -- such as cosmological phase transitions~\cite{Freese:2023fcr,Elor:2023xbz,Buckley:2024nen} and gravitational particle production during inflation~\cite{Chung:2004nh,Chung:2013rda,Chung:2013sla,Graham:2015rva,Kolb:2020fwh,Kolb:2023ydq,Amin:2022nlh,Garcia:2023qab} -- predict isocurvature power spectra that deviate from this assumption. Limits derived under the assumption of a power law cannot straightforwardly be applied to models with different power spectra.

In this work, we develop general limits on isocurvature initial conditions by using two complementary parameterizations of the isocurvature power spectrum. Such limits can be straightforwardly applied to a wide variety of new physics scenarios, removing the need for time-consuming and challenging analyses for each new modification to $\Lambda$CDM.

For the first parameterization of the power spectrum we use a delta-function in wavenumber. Although no models produce an exact delta function, this choice captures the primary characteristics of a peaked spectrum. Additionally, the delta-function power spectrum allows us to decompose the experimental limits on an arbitrary isocurvature power spectrum as a function of wavenumber. Up to $\mathcal{O}(1)$ factors, the limits on the delta-function power spectrum can be applied to an arbitrary extended power spectrum for each wavenumber.\footnote{
We note that the extrapolation of limits from a delta function to an extended power spectrum is unreliable if the power spectrum amplitude is degenerate with other parameters in the cosmological fit.}
A similar approach has been used for constraints on the curvature power spectrum and the density of primordial black holes (see, for example, \cite{PhysRevD.100.063521,Carr:2017jsz}).

To complement the delta-function approach, we also set limits on a broken power law spectrum behaving as $\propto k^3$ for small $k$ and $\propto k^0$ for large $k$. This form of power spectrum is generic to many isocurvature production mechanisms operating during inflation, with the break in the power law corresponding to the horizon at the start of isocurvature production. 
For example, this spectrum can arise from inflationary non-thermal cosmological phase transitions (see e.g., Ref.~\cite{Buckley:2024nen,Barir:2022kzo}).

At large scales ($k \lesssim 0.1\,\textrm{Mpc}^{-1}$), constraints for both power spectrum models are driven primarily by measurements of the CMB angular power spectrum and Baryon Acoustic Oscillations (BAO). Future observations of large-scale structure \cite{Chung:2023syw} and 21-cm measurements \cite{deKruijf:2024voc, Sekiguchi:2013lma} are anticipated to serve as powerful tools for probing dark matter isocurvature. On galaxy scales, the matter power spectrum can be probed using Ly-$\alpha$ forest data, leading to strong constraints on isocurvature for $k \sim 1\, \textrm{Mpc}^{-1}$. 

To set limits in these regimes, we simulate the CMB and matter power spectrum using \texttt{CLASS} code~\cite{Blas:2011rf}, modified to incorporate our choices for the isocurvature power spectra. To derive constraints from large scale structure and the CMB multipoles, we perform Markov Chain Monte Carlo (MCMC) analyses with data from the CMB, BAO, and Lyman-$\alpha$ (Ly-$\alpha$) forest across a range of scales from $\sim 10^{-4}$ to $\sim 1$~Mpc$^{-1}$. We employ an updated compressed likelihood \cite{McDonald:1999dt, Pedersen:2019ieb, Pedersen:2020kaw, Pedersen:2022anu} for the Ly-$\alpha$ data, which provides the most stringent constraint at this scale. Notably, the compressed likelihood encodes information from both the amplitude and slope of the matter power spectrum, making our constraint stronger than previous analyses that only considered the amplitude. However, these constraints can only be applied to the broken power law spectrum, as the assumption of a constant slope in the compressed likelihood is not realized in the delta function spectrum.

At even smaller scales, isocurvature perturbations induce spectral distortions in the CMB photon distribution, which appear as deviations away from a perfect blackbody spectrum. The non-observation of such distortions place further constraints on isocurvature. We calculate the isocurvature-induced amplitude of the $y$- and $\mu$-type spectral distortions, and use current bounds by COBE/FIRAS \cite{Fixsen:1996nj} to set limits for isocurvature  power spectra with wavenumbers in the range $1\, \textrm{Mpc}^{-1}\lesssim k \lesssim 10^4 \, \rm Mpc^{-1}$. 

By combining these analyses, we map out constraints on both the delta function and broken power law parametrization of the isocurvature power spectrum as a function of wavenumber, from $10^{-4}\,\textrm{Mpc}^{-1}$ to $10^{4}\,\textrm{Mpc}^{-1}$. In Section~\ref{sec:parameterizations}, we describe the types of isocurvature modes considered and the two generic power spectrum parameterizations. In Section~\ref{sec:CMB+BAO}, we present constraints from the CMB angular power spectrum and BAO data. Constraints from the Ly-$\alpha$ forest are discussed in Section~\ref{sec:Lyalpha}, and CMB spectral distortions in Section~\ref{sec:spectral_distortion}. We conclude in Section~\ref{sec:conclusions}.

\section{General Parameterizations of Isocurvature Power Spectrum} \label{sec:parameterizations}

The effects of isocurvature on cosmological observables depend on the isocurvature power spectrum and the type of isocurvature mode present. Both of these quantities depend on detailed assumptions made within a specific cosmological model containing physics beyond $\Lambda$CDM. The goal of this study is to set constraints on isocurvature that can be applied to a variety of models. 

The power spectrum quantifies the statistics of the initial conditions of the isocurvature mode and itself depends on the isocurvature production mechanism. To obtain model-independent limits, we parameterize the isocurvature power spectrum in two generic ways.

The first parameterization is a delta-function in wavenumber:
\begin{equation}\label{eq:peaked power spectrum}
    P_{\rm iso}(k) = A_{\rm iso} \delta(\ln{k}-\ln{k_0}).
\end{equation}
This power spectrum allows us to isolate the effect of isocurvature on relevant observables at each wavenumber. 
Second, we study a realistic (but still generic) parameterization of the power spectrum given by a broken power law:
\begin{equation}\label{eq:smooth power spectrum}
    P_{\rm iso}(k) = A_{\rm iso}\left\{
    \begin{array}{cc}
        (k/k_0)^3 & \quad k\leq k_0\\
        1 & \quad k>k_0
    \end{array}\right..
\end{equation}
The features of this power spectrum can arise from many models. Length scales over which there are no correlations in the isocurvature modes have a $\propto k^3$ dependence in the power spectrum. Such vanishing correlations are expected for modes that are outside the horizon for the entirety of the isocurvature production, independent of the mechanism. Isocurvature production mechanisms operating during inflation (e.g., non-thermal cosmological phase transitions~\cite{Buckley:2024nen}) can lead to an approximately scale-invariant $k^0$ power spectrum for wavenumbers that were inside the horizon during the production of isocurvature. 

For this study, we focus on isocurvature effects only and so we assume there is no new physics contribution in adiabatic modes. Therefore, we adopt the standard $\Lambda$CDM adiabatic power spectrum:
\begin{eqnarray}\label{eq:P_ab}
    P_{\rm ad}(k)=A_s \left(\frac{k}{k_{\rm pivot}}\right)^{n_s-1},
\end{eqnarray}
where the free parameters  $A_s$ and $n_s$ are the scalar amplitude and spectral index. $k_{\rm pivot}$ is the pivot scale, conventionally defined as $k_{\rm pivot} \equiv 0.05\,{\rm Mpc}^{-1}$ \cite{Planck:2018vyg}.

The initial conditions for \texttt{CLASS} are defined at a sufficiently early time $\tau_{\rm ini}$ such that $k\tau_{\rm ini}\ll1$ for all wavenumbers of interest. For the isocurvature initial conditions, we must specify which species deviates from the adiabatic condition in the limit $k \tau_{\rm ini} \to 0$ (we assume that only a single species has an initial isocurvature mode). For this work, we consider each of four different isocurvature modes: cold dark matter density isocurvature (CDI), baryon density isocurvature (BDI), neutrino density isocurvature (NDI), and free-streaming dark radiation density isocurvature (DRDI).\footnote{There is another possible isocurvature mode: neutrino velocity isocurvature. We do not study this type of isocurvature as it lacks a well-motivated generation mechanism.} 
For each of these modes, we use the standard initial conditions~\cite{Bucher:1999re,Wands:2000dp,Gordon:2000hv,Lyth:2002my,Malik:2004tf,Wands:2007bd,Ghosh:2021axu} which are defined by setting the density perturbation $\delta\equiv (\rho -\bar\rho)/\bar\rho$ for the relevant species to unity at leading order in $k\tau_{\rm ini}$.
The rest of the stress-energy and metric perturbations are specified by the evolution equations in the super-horizon limit, while requiring that there is initially no curvature in the isocurvature mode.

The full set of initial conditions for these four isocurvature modes (including the adiabatic terms), in the synchronous gauge, are shown in Appendix~\ref{app:iso_ic}. Here we list the leading term in the density perturbation of photons $\delta_{\gamma}$ in the limit of $k\tau\to0$:
\begin{equation}\begin{aligned}\label{eq:ini_iso_gamma}
    \delta_{\gamma}^{\rm CDI/BI}& =-\frac{2}{3}\frac{\Omega_{c/b}}{\Omega_m}\omega_{m}\tau, \\
    \delta_{\gamma}^{\rm NDI/DRDI} & =\frac{-R_{\nu/\rm dr}}{1-R_{\nu/\rm dr}},
    \end{aligned}
\end{equation}
where 
\begin{equation}\begin{aligned}
    \omega_{m} & \equiv \sqrt{\frac{8\pi G}{3}} \frac{a(\tau_{\rm ini})\bar\rho_m(\tau_{\rm ini})}{\sqrt{\bar\rho_{r}(\tau_{\rm ini})}}, \\
    R_{\nu/\rm dr} &\equiv \frac{\bar\rho_{\nu/\rm dr}}{\bar\rho_r},
    \end{aligned}
\end{equation} 
the total matter background density is $\bar\rho_m$, the total background radiation density is $\bar\rho_r$, and the fractional matter density is $\Omega_{c/b}/\Omega_{m}$ in CDM or baryons (for the CDI and BDI isocurvature modes, respectively).

From Eq.~\eqref{eq:ini_iso_gamma} and Appendix~\ref{app:iso_ic}, we can see that initial conditions for photons and metric perturbations in the CDI and BDI modes are the same up to a simple scaling of $\Omega_{i}/\Omega_m$ with $i\in\{c,b\}$. Given these initial conditions as well as the forms of the metric and photon evolution equations, the evolved photon perturbation continues to be proportional to $\Omega_{i}/\Omega_m$.  Since the CMB angular power spectra come from the two-point functions of photon density perturbations, the isocurvature contribution scales with $(\Omega_i/\Omega_m)^2A_{\rm iso}$ (see Eq.~\eqref{eq:C_l moments}). We have checked using \texttt{CLASS} that CDI and BDI provide identical CMB angular power spectra for the same $P_{\rm iso}$ when rescaling $A_{\rm iso}$ by $(\Omega_i/\Omega_m)^2$, with $i=c,b$. Therefore, for CDI and BDI we will present constraints on $(\Omega_i/\Omega_m)^2 A_{\rm iso}$. 

Another feature of note for CDI and BDI modes is that photon perturbations grow linearly with conformal time $\tau$ for super-horizon modes (see Eq.~\eqref{eq:ini_iso_gamma}). As a result, the size of photon perturbations for a given $k$ mode scales inversely with $k$ at the time of horizon entry (when $k\tau\sim 1$), with the associated suppression in the CMB observables at high $k$ (or high $\ell$).

For DRDI, the observables depend not only on the energy density of DR (usually parametrized as the effective number of neutrino species $\Delta N_{\rm eff}\equiv 3.044\times\bar\rho_{\rm dr}/\bar\rho_{\nu} \propto R_{\rm dr}$) through the  initial photon perturbation, which is proportional to $R_{\rm dr}/(1-R_{\rm dr})$ (see Eq.~\eqref{eq:ini_iso_gamma}), but also through changes to the background radiation energy density and thus the Hubble parameter. However, for sufficiently small $\Delta N_{\rm eff}$ (equivalently, small $R_{\rm dr}$), the modification to the total radiation density is negligible and the only effect on observables from DRDI is from isocurvature. For these small values of $R_{\rm dr}$, the initial photon perturbation is proportional to $R_{\rm dr}$ and (using the same argument as in the CDI and BDI cases) this proportionality also holds for the evolved photon perturbation variables. Therefore, in this case the effect of DRDI on observables scales with $R_{\rm dr}^2 A_{\rm iso}$. For this reason, we assume $R_{\rm dr} \ll 1$ and show constraints on DRDI in terms of $R_{\rm dr}^2 A_{\rm iso}$.

\section{Constraints from CMB+BAO}
\label{sec:CMB+BAO}

\begin{figure*}
    \centering
\includegraphics[width=\columnwidth]{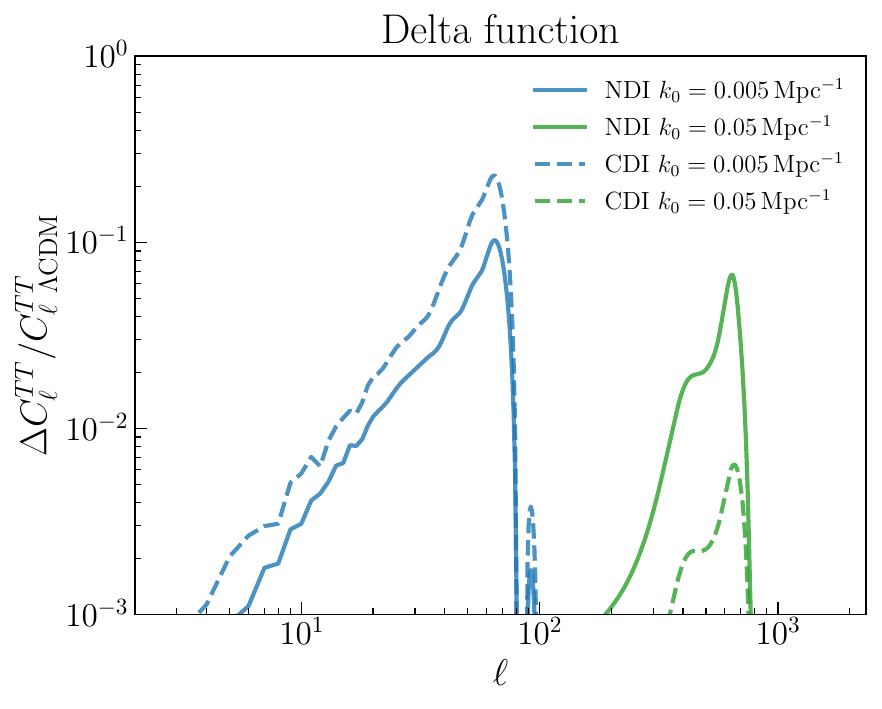}
\includegraphics[width=\columnwidth]
{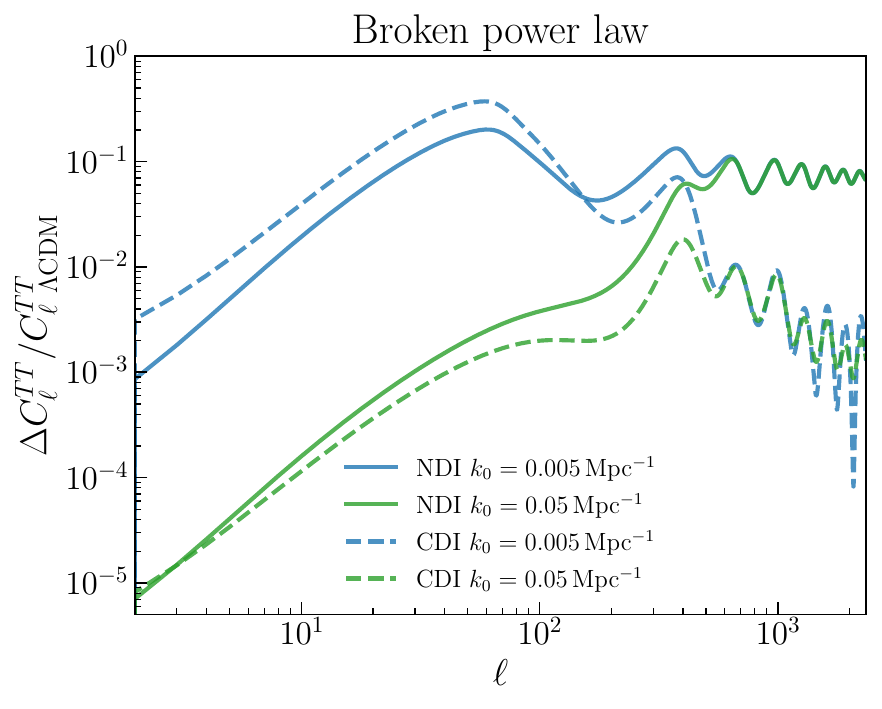}
    \caption{The fractional difference of the CMB angular power spectra $\Delta C_\ell^{TT}$ with respect to the case of $\Lambda$CDM for CDI (dashed) and NDI (solid) modes with two general forms of isocurvature power spectrum: a delta function (left, see Eq.~\eqref{eq:box_function}) and a broken power law (right, see Eq.~\eqref{eq:smooth power spectrum}). Here we choose $A_{\rm iso}=2.1\times 10^{-9}$ and two values of $k_0$: $0.005\,\textrm{Mpc}^{-1}$ (blue) and $0.05\,\textrm{Mpc}^{-1}$(green).}
    \label{fig:CTT_iso}
\end{figure*}

In this section we set constraints on $P_{\rm iso}$ at large scales, using CMB angular power spectra and BAO measurements. For the CMB, we used the angular power spectrum of the temperature (T) and polarization (E) anisotropies' two-point correlation functions. The angular power spectrum of the temperature-temperature correlation function $C^{TT}_\ell$ is
\begin{equation}\label{eq:C_l moments}
\begin{split}
    C^{TT}_\ell = 4\pi\int d(\ln{k}) \left(P_{\rm ad}(k) |\Delta_\ell^{\rm ad}(k)|^2 + P_{\rm iso}(k) |\Delta_\ell^{\rm iso}(k)|^2\right).
\end{split}
\end{equation}
In this expression, $P_{\rm ad}(k)$ and $P_{\rm iso}(k)$ are the adiabatic and isocurvature power spectra respectively, and $\Delta_\ell^A(k)$ is the photon transfer function for $A=\{{\rm ad, iso}\}$. The form of the transfer function for $A={\rm iso}$ depends on the type of isocurvature mode present. Similar expressions hold for the angular power spectrum of EE and the cross-correlations between T and E. 

For a specified isocurvature power spectrum, we calculate Eq.~\eqref{eq:C_l moments} using the Boltzmann solver \texttt{CLASS} \cite{Blas:2011rf}. Since \texttt{CLASS} requires a finite power spectrum, we regularize the peaked power spectrum Eq.~\eqref{eq:peaked power spectrum} using the parameterization 
\begin{equation}\label{eq:box_function}
    P_{\rm iso}(k) = \left\{
    \begin{array}{cc}
        A_{\rm iso}/(2\epsilon) & \quad \left|\ln{\left(k/k_0\right)}\right|\leq\epsilon\\
        0 & \quad \left|\ln{\left(k/k_0\right)}\right|>\epsilon
    \end{array}\right. ,
\end{equation}
where $\epsilon$ is a constant that determines the width (in $\log k$) of the peak. 
This expression approaches Eq.~\eqref{eq:peaked power spectrum} in the limit where $\epsilon \to 0$. In practice, it is not feasible to take $\epsilon$ to be arbitrarily small in numerical simulations.
In this study, we choose the relatively small value $\epsilon=0.05\ln(10)$. We have verified that our constraints are insensitive to the actual value of $\epsilon$ in the range $0.01\lesssim\epsilon\lesssim 0.1$, with at most a $\sim 20\%$ difference in the results. 

The simulated fractional difference of CMB temperature anisotropy angular power spectra $\Delta C_{\ell}^{TT}$ with respect to the $\Lambda$CDM case  for different isocurvature modes with our two general parameterizations of $P_{\rm iso}$ are shown in Figure~\ref{fig:CTT_iso}. We choose to show results from CDI and NDI as representative for isocurvature modes in matter and radiation respectively. 

For the delta-function power spectrum peaking at $k_0$ (Figure~\ref{fig:CTT_iso}, left panel), we see the fractional difference between the CMB power spectra with isocurvature and $\Lambda$CDM exhibits peaks at the spherical harmonic mode $\ell_0$ corresponding to $k_0$ times the comoving distance to the surface of last scattering ($d_{\rm SLS} \sim 10^4$~Mpc): $\ell_0 = k_0d_{\rm SLS}$.
We note that a delta function form for $P_{\rm iso}(k)$ does not lead to a delta function in $\Delta C^{TT}_{\ell}$ because the transfer function $\Delta_{\ell}(k_0)$ has non-zero values for a range of $\ell$ (see Eq.~\eqref{eq:C_l moments}). Moreover, we see that for fixed $A_{\rm iso}$, the ratio of height of the peaks for different $k_0$ is more dramatic for CDI than for NDI. This is because the CDI transfer function has additional $\ell$ dependence due to the time dependence in CDI initial conditions (see Eq.~\eqref{eq:ini_iso_gamma} and discussions below it).
As a result, the peak for CDI at high $\ell$ is more suppressed compared to that of NDI.

The right panel of Figure~\ref{fig:CTT_iso} shows the CMB power spectra for the CDI and NDI isocurvature modes, assuming a broken power law $P_{\rm iso}$ (see Eq.~\eqref{eq:smooth power spectrum}). We can see that for NDI, curves with different $k_0$ approach the same plateau (with some oscillations) for large $\ell$. This is because the $P_{\rm iso}$ has same constant value for $k>k_0$ for fixed $A_{\rm iso}$, regardless of $k_0$.  Each curve deviates from the common plateau when $\ell<\ell_0$ (corresponding to $k<k_0$), and asymptotes to $\propto\ell^3$ reflecting the $\propto k^3$ feature in $P_{\rm iso}$. The CDI cases, on the other hand, do not have a plateau for large $\ell$ even if $P_{\rm iso}$ is constant. This is again due to the fact that $\Delta_{\ell}^{\rm CDI}$ has additional scale dependence that leads to a suppression at high $\ell$. The oscillations in these curves reflect the phase shift between isocurvature and adiabatic modes~\cite{Savelainen:2013iwa,Baumann:2015rya}.
\subsection{Data Sets and Methodology}
To assess the cosmological constraints on the isocurvature power spectrum, we conduct a comprehensive likelihood analysis using a combination of cosmological datasets. From the relevant two-point observables computed using \texttt{CLASS}, we employ \texttt{MontePython} \cite{Brinckmann:2018cvx} for Markov Chain Monte Carlo (MCMC) exploration of the parameter space constraints from data. Our analysis incorporates the following experimental data:
\begin{itemize}
\item CMB: Planck 2018 temperature and polarization spectra, including low-$\ell$ TT, EE measurements and high-$\ell$ TTTEEE, as well as gravitational lensing reconstructions \cite{Planck:2018vyg}.
\item BAO: Baryon acoustic oscillation measurements from the Six-degree Field Galaxy Survey (6dFGS) \cite{Beutler:2011hx}, the Sloan Digital Sky Survey (SDSS) Data Release 7 Main Galaxy Sample (MGS) \cite{Ross:2014qpa}, and the LOWZ galaxy sample from BOSS DR12 \cite{BOSS:2016wmc}.
\end{itemize}

In each of our analyses, we assume the standard adiabatic modes for all species (with the standard power spectrum Eq.~\eqref{eq:P_ab}), plus one isocurvature mode for a single species. Both of our two general parametrizations of $P_{\rm iso}$ depend on two parameters: $A_{\rm iso}$ and $k_0$. For each isocurvature mode, we fix values of $k_0$ and run an MCMC analysis to get the constraint on  $A_{\rm iso}$ at 95\% confidence level (CL). Our MCMC analysis therefore has seven scanning parameters: $A_{\rm iso}$ and the six standard $\Lambda$CDM parameters $\{\omega_\mathrm{b}, \omega_\mathrm{c}, H_0, \log_{10}(10^{10} A_\mathrm{s}), n_\mathrm{s}, \tau\}$.  All cosmological parameters are assigned flat priors. Following the convention adopted by the Planck Collaboration \cite{Planck:2018vyg}, we model free-streaming neutrinos as two massless species and one massive species with $m_\nu = 0.06$ eV. 

\subsection{Results}

The 95\% CL constraints from CMB data on $A_{\rm iso}$ for different $k_0$ and isocurvature modes are summarized in Figure~\ref{fig:final limits}. For NDI, we place constraints directly on $A_{\rm iso}$, while we rescale the constraint on $A_{\rm iso}$ for CDI/BDI with $(\Omega_{c/b}/\Omega_m)^2$ in order to show the constraints on different species in the same plot. For DRDI, we choose to run the analysis for a single value of $\Delta N_{\rm eff}=0.01$ and extrapolate limits on the combination of $R_{\rm dr}^2 A_{\rm iso}$ from this. We have checked using \texttt{CLASS} simulation that this result can be applied generally to DRDI for $R_{\rm dr}\ll 1$.

For both the delta-function and broken power law parametrization of $P_{\rm iso}$, the MCMC analysis chain is used to a set a constraint on $A_{\rm iso}$ for $k_0 \in [10^{-4}, 0.2]~{\rm Mpc}^{-1}$.  As seen in the left panels of Figure~\ref{fig:final limits}, for the delta-function power spectrum the strongest constraint on $A_{\rm iso}$ occurs for $k_0$ around the CMB pivot scale $k_{\rm pivot}=0.05 \, \rm{Mpc}^{-1}$. The constraints weaken for both smaller and larger $k_0$. The strong constraints at the CMB pivot scale can be traced to the fact that the Planck data has the smallest error bars around this wavenumber. 

The broken power law spectrum results are shown in the right panels of Figure~\ref{fig:final limits}. Here, we find that the constraints on $A_{\rm iso}$ for all isocurvature modes roughly remain constant for $k_0\lesssim k_{\rm pivot}$ and scale as $k_0^3$ for  $k_0\gtrsim k_{\rm pivot}$. This behavior again arises from the strong constraining power of Planck data at $k_{\rm pivot}$. Therefore, the limit on the continuous isocurvature power spectrum is mostly set by its amplitude at this pivot scale.  For $k_0\lesssim k_{\rm pivot}$, the isocurvature spectrum (see Eq.~\eqref{eq:smooth power spectrum}) is at its plateau at the pivot scale and thus the constraint is roughly constant. For $k_0\gtrsim k_{\rm pivot}$, the size of $P_{\rm iso}(k_{\rm pivot})$ scales as $A_{\rm iso} (k_{\rm pivot}/k_0)^3$. Therefore the constraint on $A_{\rm iso}$ scales as $k_0^3$. Given this $k_0^3$ scaling, we show the MCMC results up to $k_0= 0.5 \, \rm{Mpc}^{-1}$ and then extrapolate the constraints for $k_0\geq 0.5 \, \rm{Mpc}^{-1}$ (denoted as dashed lines).

\section{Constraints from Ly-$\alpha$ Forest}
\label{sec:Lyalpha}

\begin{figure}[t]
    \centering
\includegraphics[width=\columnwidth]{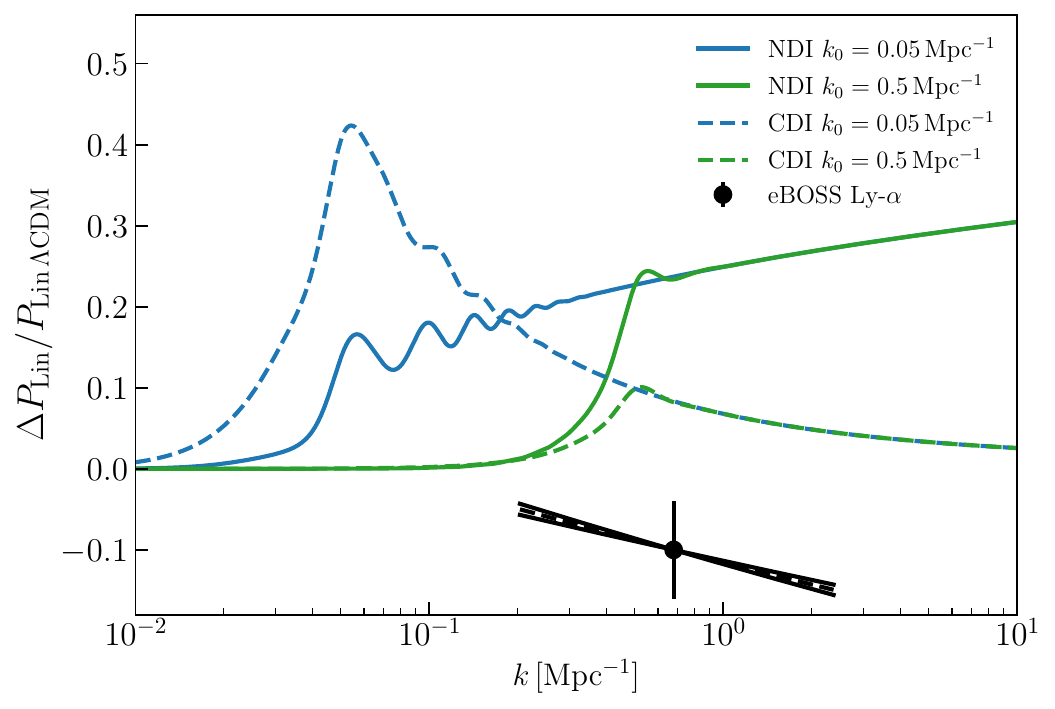}
    \caption{The fractional difference of the linear matter power spectrum $\Delta P_{\rm lin}$ between $\Lambda$CDM and the broken power law isocurvature power spectrum assuming CDI (dashed) and NDI (solid). Here we choose $A_{\rm iso}=1.05\times 10^{-8}$ and two values of $k_0$: $0.05\,\textrm{Mpc}^{-1}$ (blue) and $0.5\,\textrm{Mpc}^{-1}$(green). We also show the compressed likelihood from eBOSS Ly-$\alpha$ data~\citep{Chabanier:2018rga,Goldstein:2023gnw}.}
    \label{fig:Pk_iso}
\end{figure}

Traditional analyses of the Ly-$\alpha$ forest flux power spectrum rely on hydrodynamical simulations that scan a multi-dimensional parameter space of cosmology and astrophysical inputs. Although accurate, these analyses are computationally intensive. Instead, it has been shown \citep{McDonald:1999dt, Pedersen:2019ieb, Pedersen:2020kaw, Pedersen:2022anu} that most of the cosmological information in the Ly-$\alpha$ forest can be captured in only two parameters describing the amplitude and tilt of the linear matter power spectrum $P_{\rm lin}$ at a specific pivot redshift $z_p$ and wavenumber $k_p$:
\begin{equation}
\Delta_L^2 \,\equiv\, \frac{k_p^3\,P_{\mathrm{lin}}(k_p,z_p)}{2\pi^2}, 
\quad 
n_L \,\equiv\, \left.
\frac{d \ln P_{\mathrm{lin}}(k,z)}{d \ln k}\right|_{(k_p,z_p)}.
\end{equation}

Following Refs.~\citep{McDonald:1999dt, Pedersen:2019ieb, Pedersen:2020kaw, Pedersen:2022anu}, we compress the cosmological information from the Ly-$\alpha$ forest flux power spectrum into the amplitude and tilt of the linear matter power spectrum. We evaluate these parameters at a pivot redshift of $z_p = 3$ and a pivot wavenumber $k_p = 0.009~\mathrm{s\,km}^{-1}$ in velocity units, which corresponds to the comoving scale $k_p a(z_p)H(z_p) \approx 1 h \, \rm{Mpc}^{-1}$.
This procedure significantly reduces the computational burden compared to full hydrodynamical simulations. 
However as the compressed likelihood is only accurate if the matter power spectrum is approximately described by an amplitude and slope near $k_p$, this technique can only be applied when the primordial power spectrum is smooth without oscillations and sharp change in slope around this scale. 
We therefore set Ly-$\alpha$ forest constraint on isocurvature only when assuming the broken power law. The delta function power spectrum is not constrained by this analysis.

We use the eBOSS (SDSS DR14) Ly-$\alpha$ forest flux power spectrum measurements \citep{Chabanier:2018rga}, adopting the 2D Gaussian likelihood presented in Ref.~\citep{Goldstein:2023gnw}. 
This likelihood is marginalized over astrophysical uncertainties in the flux power modeling (e.g., thermal history and feedback processes), and its use has been validated for a wide range of cosmological models, including non-standard scenarios \citep{Pedersen:2022anu}. 
The best-fit parameters for the compressed likelihood are $\bar{\Delta}_{L}^{2}=0.310$, $\sigma_{\Delta_L^2}=0.020$, $\bar{n}_{L} = -2.340$,  $\sigma_{n_L} = 0.006$ and $\rho = 0.512$. Here $\rho$ is the correlation coefficient between $\Delta_L^2$ and $n_L$. Examples of the fractional change in the matter power spectrum relative to the $\Lambda$CDM result for both NDI and CDI isocurvature are shown in Figure~\ref{fig:Pk_iso}. We show the constraints from the compressed likelihood on the amplitude and slope of the linear power spectrum across the range of applicable wavenumbers $[0.3,3.5]~h\,{\rm Mpc}^{-1}$ . As can be seen, for both CDI and NDI cases, if the transition in the broken power law occurs away from pivot wavenumber $k_p$, the resulting linear matter power spectrum is smooth through the region where the compressed likelihood is used to set bounds. Transitions near $k_p$ lead to more rapidly changing slopes in the power spectrum, making the limits extracted from the compressed likelihood less robust.

In particular, the compressed Ly-$\alpha$ likelihood allows one to isolate the power-spectrum amplitude and tilt, mitigating biases that might arise when projecting higher-dimensional parameter constraints. 
This approach ensures consistency across different cosmic epochs and scales, thereby improving our overall cosmological parameter inferences.
The compressed likelihood is added to that of the CMB, BAO, and SNe measurements described in Section~\ref{sec:CMB+BAO}, again using \texttt{MontePython} to explore the parameter space via MCMC.
In this combination, we evaluate the log-likelihood of each dataset independently and derive joint posteriors in parameter space. 

The resulting constraints on the isocurvature model parameters are shown in Figure~\ref{fig:final limits}. The joint constraints with CMB+BAO and Ly-$\alpha$ are approximately the same as the CMB+BAO ones for $k_0\lesssim 0.2\,\textrm{Mpc}^{-1}$. This occurs because the plateau of the power spectrum extends to scales that the CMB is sensitive to and thus constraints are dominated by CMB+BAO. For $k_0\gtrsim 1\,\textrm{Mpc}^{-1}$, CMB constraints weaken, while Ly-$\alpha$ constraints are stronger as the slope of $P_{\rm lin}$ deviates from the preferred value of the compressed likelihood near the pivot scale $k_p$ because $P_{\rm iso}\propto (k/k_0)^3$. In this regime, the constraints scale as $k^3_0$, and we extrapolated the results for $k_0\geq 3\,\textrm{Mpc}^{-1}$ with this scaling (denoted as dashed lines). There is an intermediate region around $k_0\sim1\, h\, \textrm{Mpc}^{-1}$ (corresponding to $k_p$) where the constraints exhibit a sharp transition. This feature is due to the rapid change in slope of the isocurvature power spectrum and the resulting linear matter power spectrum around $k_p$. Since the slope of $P_{\rm lin}$ quickly changes near $k_p$ for these values of $k_0$, we again note that the constraints in this regime are less robust due to limitations in the compressed likelihood approach.

\section{Constraints from CMB Spectral Distortions}
\label{sec:spectral_distortion}

\begin{figure*}
\centering
\includegraphics[width=2\columnwidth]{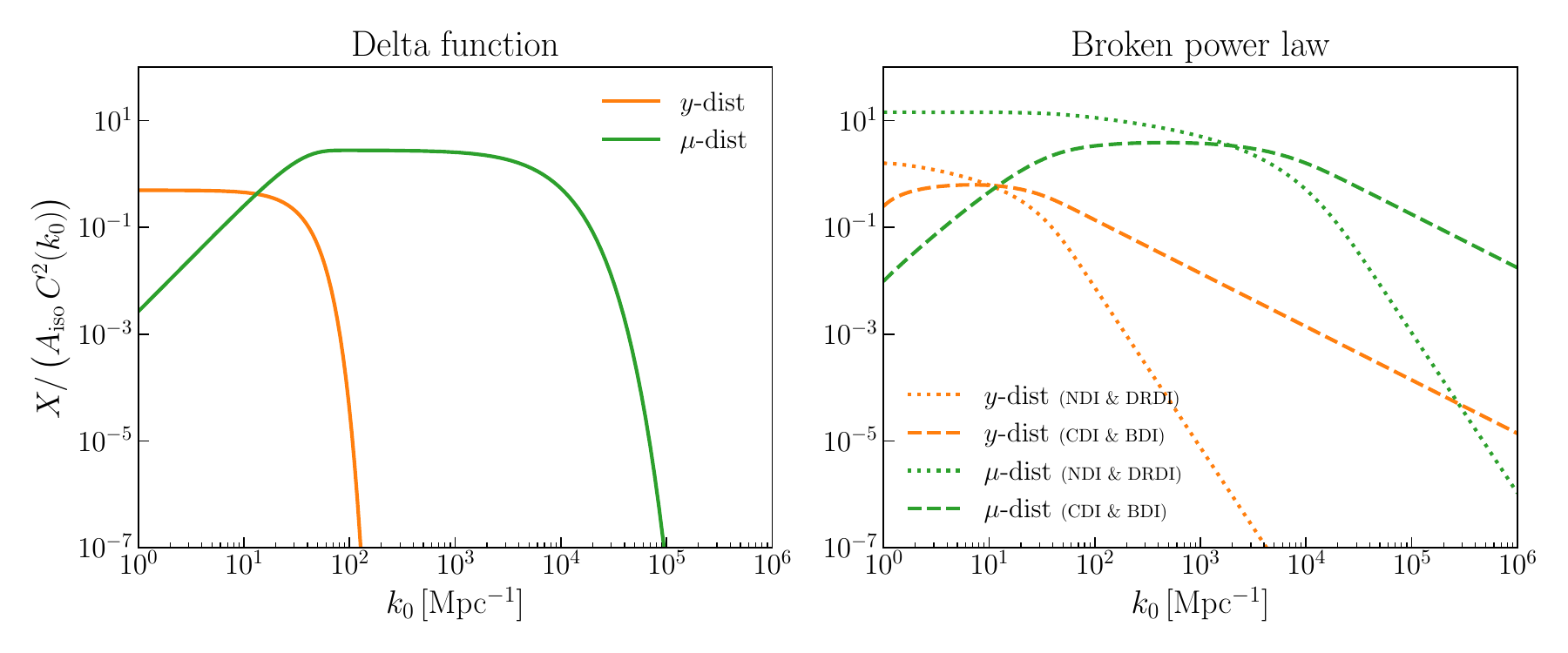}
\caption{Spectral distortions of type $X \in \{y, \mu \}$ normalized to $A_{\rm iso}C^2(k_0)$ for the delta function power spectrum (left) and the broken power law spectrum (right). For the delta function power spectrum, each type of isocurvature mode leads to spectral distortions that are the same up to the proportionality factor $A_{\rm iso}C^2(k_0)$. We therefore show a single curve for all modes. For the broken power law spectrum, the distortions for NDI and DRDI modes only differ by the proportionality factor $A_{\rm iso}C^2(k_0)$ and are shown with dotted curves. The distortions for CDI and BDI mode as functions of $k_0$ do not have the same form and are shown with the dashed curves.}
\label{fig:spectral distortions}
\end{figure*}

In this section we consider constraints on isocurvature modes from CMB spectral distortions. These constraints apply for wavenumbers in the range $1\, \textrm{Mpc}^{-1}\lesssim k\lesssim 10^4\, \rm Mpc^{-1}$, complementary to the sensitivity range of CMB anisotropies and BAO data.

The observed CMB frequency spectrum is consistent with a blackbody distribution with zero chemical potential. Therefore, any dynamics (e.g., energy injection) that causes distortions in the CMB spectrum (that is, deviations from the blackbody distribution), will be constrained by current and future data (for earlier studies, see e.g.,~\cite{Chluba:2011hw,Dent:2012ne,Chluba:2012gq,Khatri:2012rt,Chluba:2013dna,Chluba:2013wsa,Arsenadze:2024ywr}). Depending on the time of the energy injection and the relevant processes in the photon and baryon bath that are active, the spectral distortion can be parametrized by different forms. At very early times, $z\gtrsim2 \times 10^6$, the rates of photon number-changing processes (such as double Compton scattering and Bremsstrahlung) are much larger than the Hubble expansion rate. Therefore, any kind of energy injection to the photon and baryon bath will be quickly thermalized, creating a blackbody spectrum at a different temperature with a vanishing chemical potential. This change can be fully absorbed into the measurement of CMB temperature, leaving no spectral distortions. 

For redshifts $ 5\times 10^4 \lesssim z\lesssim2 \times 10^6$, photon number-changing processes are inefficient. However, Compton scattering is still active, keeping the photon and electron plasma in thermal equilibrium. In this epoch the photon phase space density follows a Bose-Einstein distribution and exotic energy injection will in general change the temperature and generate a non-zero chemical potential. This creates a difference $\Delta f_\gamma$ between the phase space density and a blackbody spectrum:
\begin{eqnarray}\label{eq:Deltaf}
    \Delta f_\gamma (\omega,T)=\frac{1}{e^{\omega/(T+\Delta T)+\mu}-1}-f_{\gamma,0}(\omega/T)\,.
\end{eqnarray}
Here $f_{\gamma,0}(x)\equiv 1/(e^{x}-1)$ is the blackbody distribution with zero chemical potential, $\Delta T$ is the change in temperature relative to $T$, and $\mu$ is the size of the $\mu-$type spectral distortion and is proportional to the chemical potential. In our analysis, we treat both $\mu$ and $\Delta T/T$ as small perturbations and work at linear order in these parameters.  $\Delta f$ is defined such that the new phase space density has the same number density as the blackbody at temperature $T$ but a different energy density\footnote{For the dissipation of acoustic modes that we consider in this work, the change in photon number density at second order in perturbation theory can be fully absorbed into the definition of the temperature $T$ in Eq.~\eqref{eq:Deltaf}. After this redefinition, a net change in photon energy density remains.}. Using this condition the relation between $\mu$ and the change in photon energy density is 
\begin{equation}\label{eq:mu_deltarho}
      \frac{\Delta n_\gamma}{n_\gamma}\equiv\frac{\int d\omega\,\omega^2 \Delta f_{\gamma}}{\int d\omega\, \omega^2 f_{\gamma,0}}=0 \,\, 
   \Rightarrow\,\, \mu\approx1.4\frac{\Delta \rho_\gamma}{\rho_\gamma},
\end{equation}
where 
\begin{equation}
    \frac{\Delta \rho_\gamma}{\rho_\gamma}\equiv\frac{\int d\omega\, \omega^3\Delta f_{\gamma}}{\int d\omega \,\omega^3 f_{\gamma,0}}.
\end{equation}

Below $z\lesssim 5\times 10^4$, Compton scattering is inefficient at maintaining a Bose-Einstein distribution for the photons after energy injection. However, photons can still interact with electrons via Compton scattering before recombination, making the resulting photon phase space distribution a mixture of blackbody spectra with slightly different temperatures. This kind of distribution can not be described by a single blackbody with a new temperature, instead creating a spectral distortion in photon phase space called a $y$-distortion:
\begin{eqnarray}\label{eq:y_distortion}
    \Delta f_\gamma (\omega,T) =y\, \mathcal{Y}(\omega/T),
\end{eqnarray}
where
\begin{equation}
    \mathcal{Y}(x) = \left[x(1 + 2 f_{\gamma,0}(x)) - 4\right]\mathcal{G}(x)
\end{equation}
with
\begin{equation}
    \mathcal{G}(x) \equiv xf_{\gamma,0}(x)(1 + f_{\gamma,0}(x)).
\end{equation}
The coefficient $y$ in Eq.~\eqref{eq:y_distortion} denotes the size of the $y$-distortion and is related to the energy release as
\begin{equation}\label{eq:y_deltarho}
    y=\frac{1}{4}\frac{\Delta\rho_\gamma}{\rho_\gamma}.
\end{equation}

As seen in Eqs.~\eqref{eq:mu_deltarho} and \eqref{eq:y_deltarho}, the key quantity to determine the size of the spectral distortion (both $\mu$ or $y$) is the change in the photon energy density $\Delta \rho_\gamma/\rho_\gamma$. The main source of such energy injection from primordial density perturbations with $k\geq 1 \rm \, Mpc^{-1}$ is the dissipation of acoustic modes through photon diffusion~\cite{Chluba:2013dna}. This process generates spectral distortions when $z\gtrsim z_*=1100$. As a result, in our calculations we can assume that photons and baryons are always tightly coupled.
With this assumption, the energy injection rate from acoustic modes $Q_{\rm ac}$  is given by~\cite{Chluba:2013dna}
\begin{equation}\label{eq:heating_rate}
    \frac{1}{a^4\rho_\gamma} \frac{d(a^4 Q_{\rm ac})}{dz} = \frac{64}{15}\frac{1}{\kappa H(1+z)}\int d(\ln{k}) k^2 P_{\rm iso}(k) (\Theta_1^{\rm iso})^2,
\end{equation}
where $\kappa = \sigma_T n_e$ is the rate of Thomson scattering and $\Theta_1^{\rm iso}$ is the dipole moment of the photon temperature perturbation in the isocurvature mode.  As the size of isocurvature power spectra  we consider are much larger than that of curvature $P_{\rm iso}\gg P_{\rm ad}$, we only keep the contribution from isocurvature perturbations in Eq.~\eqref{eq:heating_rate}.  

For the dissipation of an acoustic mode with wavenumber $k$, the energy injection rate is non-negligible only when the diffusion scale $k_D(z)$ is of order $k$. For $z\gtrsim z_{\rm eq}$, baryon loading is small as $\rho_{b}\ll \rho_\gamma$ and the diffusion scale can be approximated by
\begin{equation}\label{eq:k_D2}
    \partial_z k_D^{-2} \approx -\frac{8}{45}\frac{1+z}{H\kappa}.
\end{equation}
To find $k_D^{-2}$ as a function of redshift $z$, we integrate Eq.~\eqref{eq:k_D2} from $z=\infty$ to $z$ and obtain the following analytic result which is valid for $z\gtrsim z_{\rm eq}$:
\begin{equation}\label{eq:diffusion scale}
\begin{split}
    k_D^{-2} = & \left(0.35\, \mathrm{Mpc}\right)^2\times \left(\frac{0.143}{\Omega_m h^2}\right)^{1/2}\left(\frac{0.022}{\Omega_b h^2}\right)\left(\frac{0.88}{1-Y_p/2}\right)\\
    \times & \left(\frac{1+z}{10^4}\right)^{-5/2} f_D\left(\frac{1+z_{\rm eq}}{1+z}\right) \,,
\end{split}
\end{equation}
where $z_{\rm eq}$ is the redshift of matter-radiation equality and
\begin{equation}
    f_D(y) \equiv \frac{1}{3y^{5/2}}\left[\sqrt{1+y}\left(3y^2-4y + 8\right) - 8\right].
\end{equation} 
Therefore, the diffusion scale is well inside the horizon at all redshifts and it satisfies $k_D(z)\gtrsim 1\,\rm Mpc^{-1}$ for $z\gtrsim z_{\rm eq}$. This means that the modes we are considering ($k\gtrsim 1 \, \textrm{Mpc}^{-1}$) will only inject energy when $z\gtrsim z_{\rm eq}$, implying that we are justified in neglecting baryon loading. Under these circumstances, the photon dipole moment can be approximated for modes inside the horizon as
\begin{equation} \label{eq:Theta_1 form}
    \Theta_1^{\rm iso} \approx c_s \left[A(k)\sin{(kr_s)} - B(k)\cos{(kr_s)}\right]e^{-k^2/k_D^2} ,
\end{equation}
where $r_s = \int dt\, c_s/a$ is the sound horizon and $c_s^2 \approx 1/3$ is the sound speed.

\begin{figure*}
\centering
\includegraphics[width=2\columnwidth]{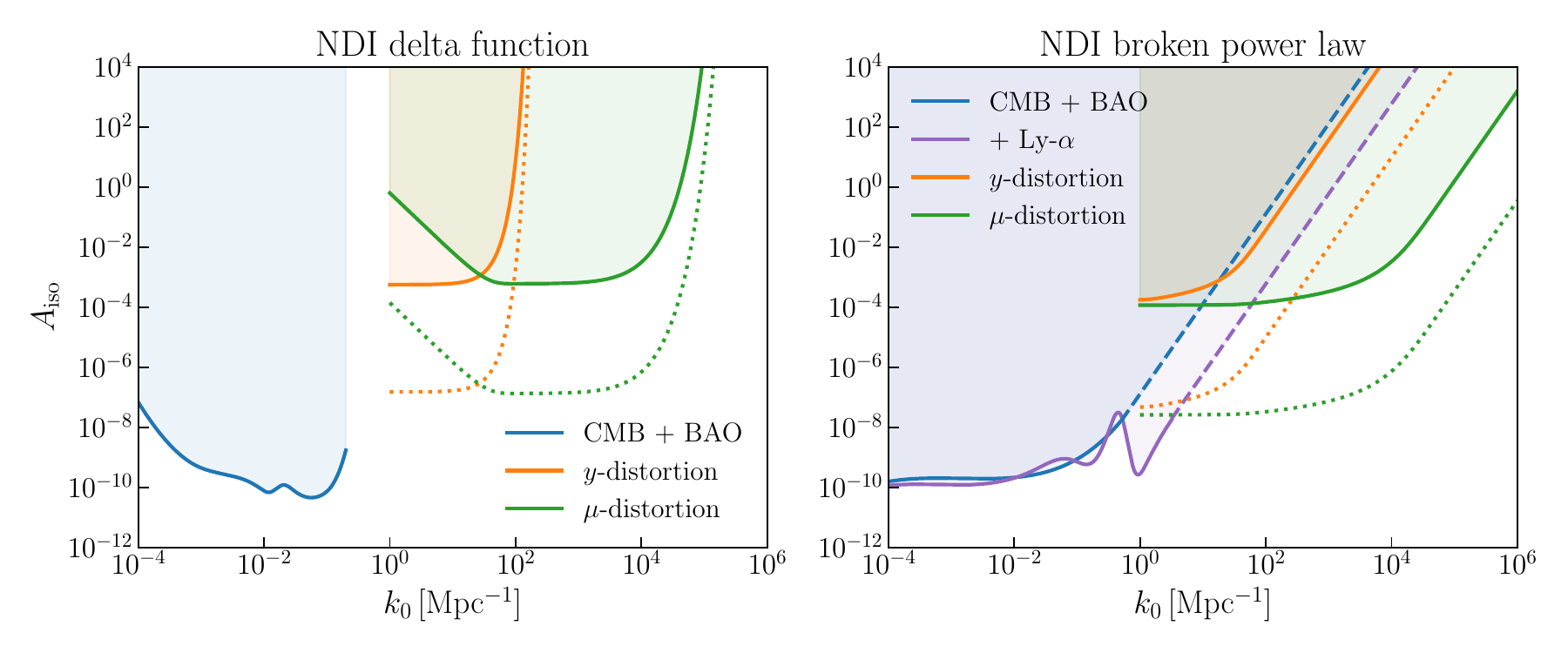}
\includegraphics[width=2\columnwidth]{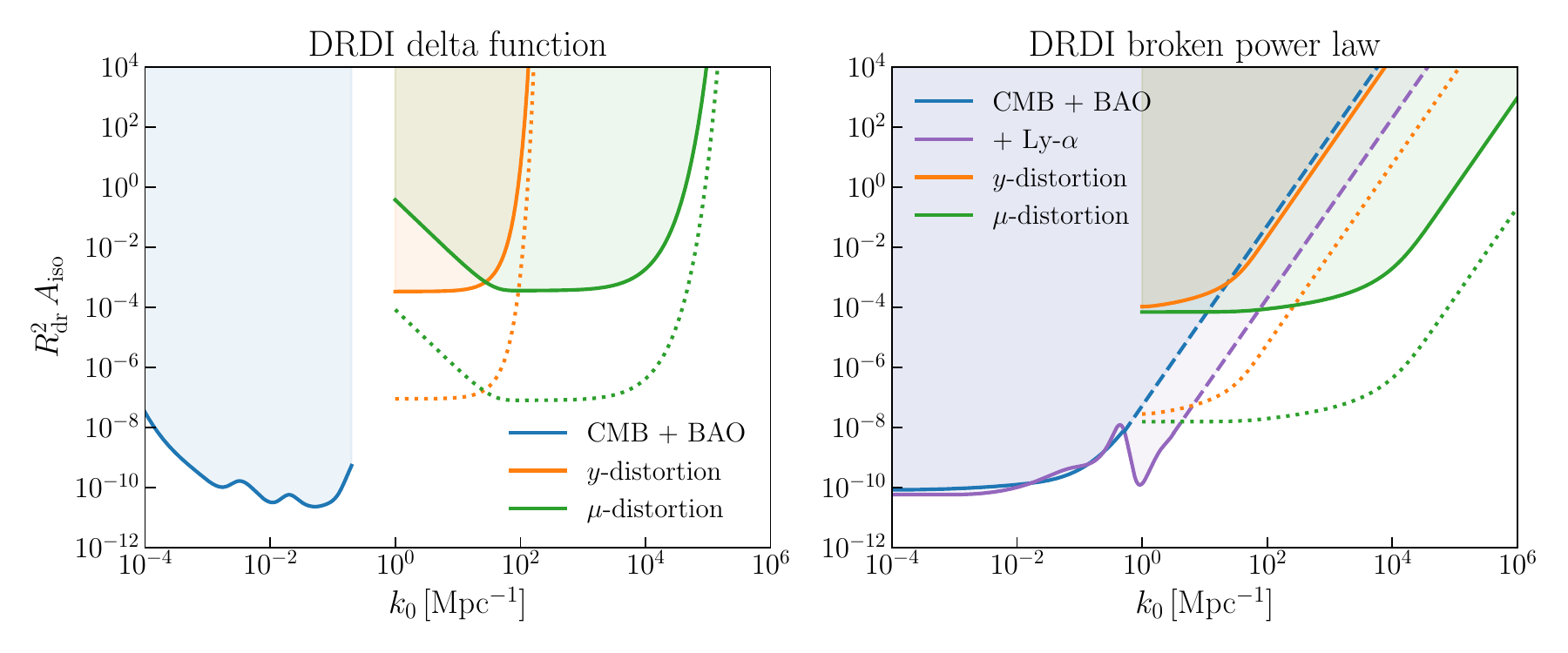}
\includegraphics[width=2\columnwidth]{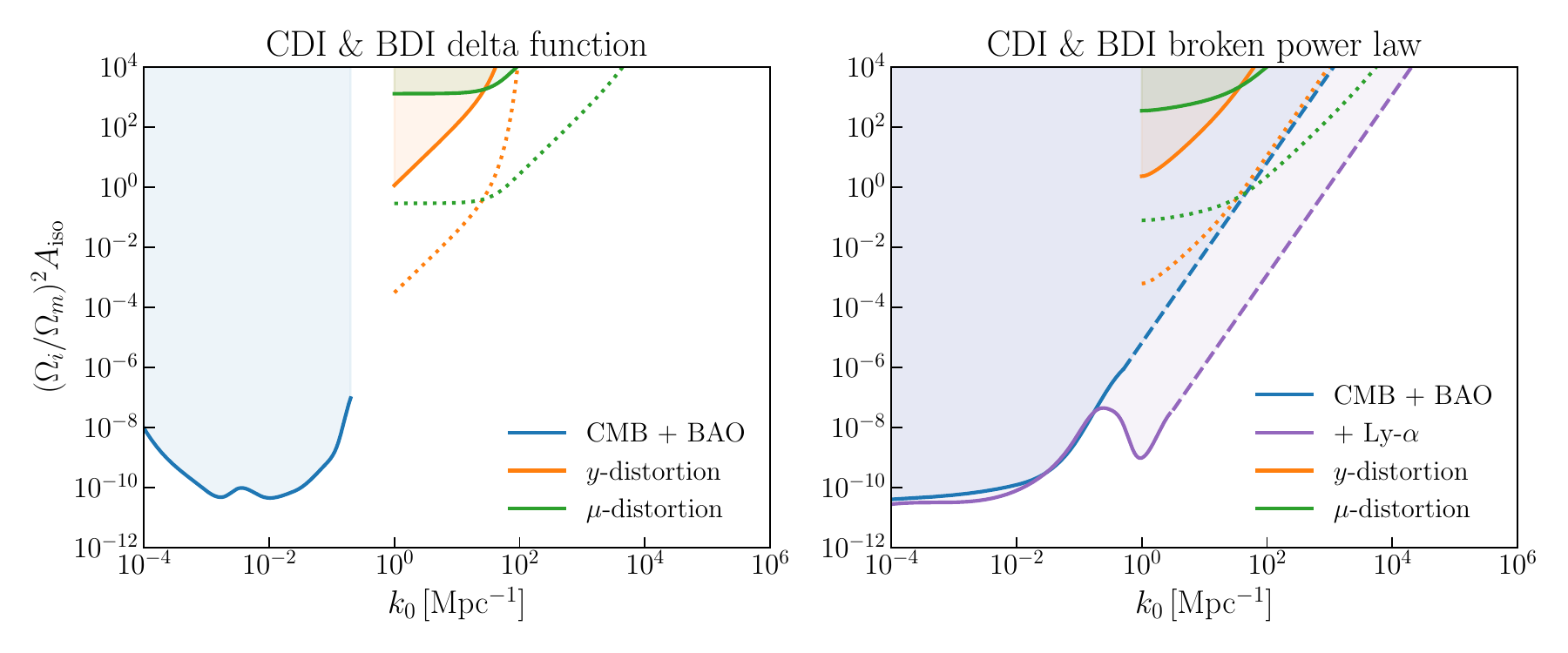}
\caption{Limits on isocurvature power spectrum amplitude as a function of $k_0$ for NDI, DRDI, CDI and BDI from various observations. The left column shows limits on the delta-function power spectrum (Eq.~\eqref{eq:peaked power spectrum}) and the right column shows limits on the broken power law spectrum (Eq.~\eqref{eq:smooth power spectrum}). 
Limits from CMB and BAO are shown in blue while joint constraints with Ly-$\alpha$ are shown in purple. The dashed lines indicate the extrapolation that is proportional to $ k_0^3$.  Limits from $y$-type and $\mu$-type spectral distortions are in orange and green, respectively. For each type of spectral distortion we show current limits (solid) from COBE/FIRAS~\cite{Fixsen:1996nj} and projected limits (dotted) for PIXIE~\cite{Kogut:2011xw}.}
\label{fig:final limits}
\end{figure*}

From Eqs.~\eqref{eq:mu_deltarho},   \eqref{eq:y_deltarho} and \eqref{eq:heating_rate}, the $\mu$ and $y$ distortions in the CMB today can be calculated in terms of the energy release during two different epochs\footnote{In principle, $y$ distortions can be generated from energy release at any redshift after $z_{\mu,y}$. However, the $y$ distortion from photon diffusion considered in this work contributes only before recombination. Therefore, we set the lower limit of the second integral in Eq.~\eqref{eq:mu_and_y} to $z_*$.}~\cite{Chluba:2013dna}:
\begin{equation}\label{eq:mu_and_y}
    \begin{split}
        \mu &\approx 1.4 \left.\frac{\Delta \rho_\gamma}{\rho_\gamma}\right|_\mu \approx 1.4 \int\displaylimits_{z_{\mu, y}}^\infty dz\mathcal{J}_{bb}(z)\frac{1}{a^4 \rho_\gamma}\frac{d(a^4Q_{ac})}{dz}\\
        y & \approx \frac{1}{4}\left.\frac{\Delta \rho_\gamma}{\rho_\gamma}\right|_y \approx \frac{1}{4}\int\displaylimits_{z_*}^{z_{\mu,y}} dz \mathcal{J}_{bb}(z) \frac{1}{a^4 \rho_\gamma}\frac{d(a^4Q_{ac})}{dz},
    \end{split}
\end{equation}
where $z_{\mu,y}=5\times 10^4$ and $\mathcal{J}_{bb}(z) = \exp{\left[-(z/z_\mu)^{5/2}\right]}$ which is the visibility function\footnote{A refined visibility function is considered in Ref.~\cite{Chluba:2013kua}. We have verified that the difference between the spectral distortions obtained in this work and the result from using the refined visibility function is less than $10\%$.} which erases contributions to $\mu$ distortions at redshifts $z\gtrsim z_\mu =1.98\times10^6$. Note that $\mathcal{J}_{bb}(z) \to 1$ in the redshift range over which the integral for the $y$ distortion is taken.

Substituting Eq.~\eqref{eq:heating_rate} into the expressions for $\mu$ and $y$ in Eq.~\eqref{eq:mu_and_y}, we see that $\Theta_1$ (defined in Eq.~\eqref{eq:Theta_1 form}) will oscillate with $z$ much faster than any other factor in the integrand. As a result, the integrals in Eq.~\eqref{eq:mu_and_y} can be approximated by averaging the integrands over many oscillation periods, allowing us to replace $\sin^2(kr_s)$ and $\cos^2(kr_s)$ each with $1/2$ and to omit the cross-term. After these approximations, the fractional energy density due to spectral distortions can be written as 
\begin{equation}\label{eq:fractional_distortion}
        \frac{\Delta \rho_\gamma}{\rho_\gamma} = 2 \int d(\ln{k}) C(k)^2P_{\rm iso}(k)\int dz \mathcal{J}_{bb}(z) \partial_ze^{-2k^2/k_D^2},
\end{equation}
where $C(k)^2 \equiv A(k)^2 + B(k)^2$ (with $A(k)$ and $B(k)$ defined in Eq.~\eqref{eq:Theta_1 form}). This expression can be evaluated for $\mu$ and $y$ distortion by using the appropriate limits of the $z$ integral.

As we are interested in limits on four different isocurvature modes (CDI, BDI, NDI and DRDI), we must calculate the coefficient $C(k)^2$ in Eq.~\eqref{eq:fractional_distortion} for each of those modes. 
In each case, the coefficients $A(k)$ and $B(k)$ -- and from them, $C(k)$ -- can be found by simulating the Boltzmann Einstein equations for the appropriate initial conditions and fitting Eq.~\eqref{eq:Theta_1 form} to the numerical result during the radiation domination era for modes inside the horizon. 

For NDI, the $k$-independent value of $C(k)^2$ was found in a previous numerical study \cite{Chluba:2013dna}:
\begin{equation}\label{eq:NDI_C}
    C(k)^2 \approx 0.052 \rm \quad (NDI) .
\end{equation}
For both CDI and BDI, analytic considerations \cite{Hu:1995en} imply that $\left\{A, B\right\} \propto (\Omega_i/\Omega_m)(k_{\rm eq}/k)$ where $\Omega_i$ is the density parameter for matter species $i$ and $k_{\rm eq} = 9.56\times 10^{-3} \, \rm Mpc^{-1}$ is the wavenumber that crosses the horizon at matter-radiation equality. This form for $A$ and $B$ was fit to the numerical result \cite{Chluba:2013dna} leading to
\begin{equation}\label{eq:CDIBDI_C}
    C(k)^2 \approx 0.28 \left(\frac{\Omega_i}{\Omega_m}\right)^2 \left(\frac{k_{\rm eq}}{k}\right)^2  \rm \quad (CDI/BDI).
\end{equation}

For free-streaming dark radiation, the analogous result does not exist in the literature. We used \texttt{CLASS} to extract the $A$ and $B$ coefficients for the DR isocurvature mode with different energy densities of dark radiation. We found that the coefficients are well approximated by the $k$-independent values
\begin{equation}
    \begin{split}
        A\approx & -0.28\times \frac{R_{\rm dr}}{1-R_{\rm dr}} \\
        B\approx & -0.1\times \frac{R_{\rm dr}}{1-R_{\rm dr}}
    \end{split}
\end{equation}
for $R_{\rm dr}\ll 1$. This leads to
\begin{equation}\label{eq:C dr}
    C(k)^2 = 0.088 \left(\frac{R_{\rm dr}}{1-R_{\rm dr}}\right)^2 \quad (\textrm{DRDI}\,,\,R_{\rm dr}\ll 1).
\end{equation}

From these results, we calculate the $y$ and $\mu$ distortions for each mode and power spectrum. According to Eqs.~\eqref{eq:mu_and_y} and~\eqref{eq:fractional_distortion}, $y$ distortions have the form
\begin{equation}\label{eq:y_final}
    y \approx \frac{1}{2}\int\displaylimits_{k_{\rm min}}^\infty d (\ln{k})C(k)^2P_{\rm iso}(k)\left.e^{-2k^2/k_D^2}\right|^{z_{\mu, y}}_{z_{*}},
\end{equation}
where we set a lower limit on the $k$ integral given by $k_{\rm min}=1\,\textrm{Mpc}^{-1}$. This choice ensures that the approximation of small baryon loading is satisfied and it is a conservative choice as any contributions to the integral from lower $k$ are positive.
For the $\mu$ distortions, we find
\begin{equation}\label{eq:mu_dist}
    \mu \approx 2.8 \int\displaylimits_{k_{\rm min}}^\infty d(\ln{k}) C(k)^2P_{\rm iso}(k)\int\displaylimits_{z_{\mu, y}}^\infty dz \mathcal{J}_{bb}(z) \partial_ze^{-2k^2/k_D(z)^2},
\end{equation}
which must be numerically integrated for each choice of power spectrum.

Using the delta-function power spectrum, the $k$ integrals are trivial and both spectral distortions are proportional to $C(k_0)^2A_{\rm iso}$ for all isocurvature modes. We show the $y$ and $\mu$ distortions normalized to this product in the left panel of Figure~\ref{fig:spectral distortions}.

Using the broken power law spectrum, the $k$ integral is non-trivial and the result depends on the full $k$ dependence of $C(k)^2$ for each mode. We show the $y$ and $\mu$ distortions for each case in the right panel of Figure~\ref{fig:spectral distortions}, again with the results normalized to $A_{\rm iso} C(k_0)^2$.
Since NDI and DRDI both have a $C(k)^2$ which is independent of $k$, the results for these modes are the same (shown with the dotted curves). On the other hand, CDI and BDI modes have $C(k)^2 \propto k^{-2}$. The results for these modes are shown with dashed curves.

The $2\sigma$ upper bounds on $\mu$ and $y$ from COBE/FIRAS are given by \cite{Fixsen:1996nj}
\begin{equation}
    \begin{split}
        |y| & \leq 1.5\times 10^{-5} \\
        |\mu| & \leq 9 \times 10^{-5}.
    \end{split}
\end{equation}
The projected limits for PIXIE are \cite{Kogut:2011xw}
\begin{equation}
    \begin{split}
        |y| & \leq 4\times 10^{-9} \\
        |\mu| & \leq 2 \times 10^{-8}.
    \end{split}
\end{equation}
The resulting existing and projected limits on isocurvature from spectral distortions are shown in Figure~\ref{fig:final limits}. For the delta function power spectrum (the left panels of Figure~\ref{fig:final limits}), the strongest constraints are from $y$-distortions for $1\,\textrm{Mpc}^{-1}\lesssim k_0\lesssim 40\,\textrm{Mpc}^{-1}$  and from  $\mu$-distortions for $40\,\textrm{Mpc}^{-1}\lesssim k_0\lesssim 10^4\,\textrm{Mpc}^{-1}$. Constraints weaken for $k_0\gtrsim 10^4\,\textrm{Mpc}^{-1}$ because these modes are within the photon diffusion length ($k_0\gtrsim k_D(z)$) for the whole period ($z\lesssim 10^6$) during which spectral distortions can be produced. For NDI and DRDI modes, the dominant constraints between $1\,\textrm{Mpc}^ {-1}\lesssim k_0\lesssim 10^4\,\textrm{Mpc}^{-1}$ are almost flat due to  $C(k)^2$ being independent of $k$ (see Eqs.~\eqref{eq:NDI_C} and~\eqref{eq:C dr}). In comparison, CDI and BDI results exhibit $k_0^2$ dependence as a result of the $C(k)^2\propto k^{-2}$ dependence in Eq.~\eqref{eq:CDIBDI_C}. 

For the case of broken power law (the right panels of Figure~\ref{fig:final limits}),  the constraint is always dominated by the $\mu$-distortion for NDI and DRDI, and behaves as $\propto k_0^3$ for $k_0\gtrsim 10^4\,\textrm{Mpc}^{-1}$ due to the $k^3$ part of $P_{\rm iso}$. For the broken power law constraints on CDI and BDI, the results are similar to that from delta function: the $y$-distortion dominates for $1\,\textrm{Mpc}^{-1}\lesssim k_0\lesssim 40\,\textrm{Mpc}^{-1}$ and the combined constraint shows $k_0^2$ dependence for $1\,\textrm{Mpc}^{-1}\lesssim k_0\lesssim 10^4\,\textrm{Mpc}^{-1}$. In principle, the limits would apply for $k_0<1 \,\rm Mpc^{-1}$ as well since the power spectrum still has support for $k>1 \,\rm Mpc^{-1}$. However, the CMB and Ly-$\alpha$ limits are significantly stronger for this range of $k_0$ so we do not show the constraints from spectral distortions in the regime.

\section{Conclusions}
\label{sec:conclusions}

New physics models can generate isocurvature signatures in cosmological observables in any species (CDM, baryons, neutrinos and dark radiation) with a wide variety of primordial isocurvature power spectra. Existing cosmological constraints on the isocurvature power spectrum generally assume a simple power law, limiting the applicability. In this work, we derive general constraints that can be applied to a broad class of new physics models by parametrizing the isocurvature power spectrum with two forms: a delta-function and a broken power law, and consider four types of isocurvature modes (CDI, BDI, NDI, and DRDI). 

We place constraints on the isocurvature power spectrum across a wide range of scales with data from the CMB+BAO, the Ly-$\alpha$ forest, and CMB spectral distortions. From our MCMC analysis, we find that CMB+BAO sets the strongest constraints at large scales ($k \lesssim 0.1\,\textrm{Mpc}^{-1}$), while Ly-$\alpha$ puts the most stringent constraint at $k \sim 1\,\textrm{Mpc}^{-1}$. CMB spectral distortions constrain the isocurvature spectrum at $ 1\, \textrm{Mpc}^{-1}\lesssim  k \lesssim 10^4\, \rm Mpc^{-1}$. Note that the Ly-$\alpha$ constraints assume a continuous power spectrum and so are only applied in this work to the broken power law parametrization of isocurvature.

Other observables can also place constraint on isocurvature spectrum at different scales. The presence of isocurvature can affect the Big Bang Nucleosynthesis (BBN) and the primordial abundance of light elements (see e.g., Ref.~\cite{Adshead:2020htj} for DRDI). Isocurvature perturbations can source gravitational waves as they enter the horizon and thus will be constrained by gravitational wave observations~(see Refs.~\cite{Domenech:2021and,Domenech:2023jve} for CDI). 
Large isocurvature perturbations can also form primordial black holes~(For CDI, see Ref.~\cite{Passaglia:2021jla}).
We leave the study of general constraints on different isocurvature modes from these effects to future work.

\section*{Acknowledgements}

We thank Subhajit Ghosh and Soubhik Kumar for useful discussions. MRB, NF and MJW are supported by DOE grant DOE-SC0010008. PD is supported by the National Natural Science Foundation of China (Grants No. T2388102).

\newpage
\appendix
\section{Adiabatic and Isocurvature initial conditions}\label{app:iso_ic}
In this section, we will present the full set of initial conditions for perturbations in adiabatic and four isocurvature (CDI, BDI, NDI and DRDI) modes in the synchronous gauge and in terms of the conformal time $\tau$ (see~\cite{Ma:1995ey,Bucher:1999re,Ghosh:2021axu}). Other than DRDI, we assume no dark radiation present in these modes. In the synchronous gauge, the perturbed FRW metric is written as
\begin{equation}
    ds^2 = a(\tau)^2 \left[-d\tau^2 + (\delta_{ij} + h_{ij})dx^idx^j\right].
\end{equation}
The metric perturbation $h_{ij}$ can be written in Fourier space as
\begin{equation}
   h_{ij}(\bs{k}, \tau) = \left[\hat{k}_i\hat{k}_jh(\bk, \tau) + \left(\hat{k}_i\hat{k}_j - \frac{1}{3}\delta_{ij}\right)6\eta(\bk, \tau)\right],
\end{equation}
where $h$ and $\eta$ denote the trace and traceless longitudinal part of $h_{ij}$ respectively. 

 We will then show initial conditions for metric perturbations $h,\eta$, as well as first three moments of density perturbations of each species (denoted as $\delta,\theta,\sigma$).

\subsection{Adiabatic modes}
The set of adiabatic modes are given by
\begin{eqnarray}
    &&\eta^{\rm ad}= 1 - \frac{5+4R_{\nu}}{12(15+4R_{\nu})}(k\tau)^2 \nonumber\\
    &&h^{\rm ad}=\frac{1}{2}k^2\tau^2\nonumber\\
    &&\delta_\gamma^{\rm ad}=-\frac{1}{3}k^2\tau^2\nonumber\\
    &&\delta_c^{\rm ad}=\delta_b^{\rm ad}=\frac{3}{4}\delta_\gamma^{\rm ad}=\frac{3}{4}\delta_\nu^{\rm ad}\nonumber\\
    &&\theta_c^{\rm ad}=0\nonumber\\
    &&\theta_\gamma^{\rm ad}=\theta_b^{\rm ad}=-\frac{1}{36} k^4\tau^3\nonumber\\
    &&\theta_{\nu}^{\rm ad}=\frac{23+4R_{\nu}}{15+4R_{\nu}}\theta_\gamma^{\rm ad}\nonumber\\
    &&\sigma_{\nu}^{\rm ad}=\frac{2}{3(15+4R_{\nu})}k^2\tau^2.
\end{eqnarray}

\subsection{CDI and BDI modes}
The set of CDI and BDI modes are given by
\begin{eqnarray}
    &&\eta^{\rm CDI}= -\frac{1}{6}\frac{\Omega_c}{\Omega_m}\omega_m\tau+\frac{1}{16}\frac{\Omega_c}{\Omega_m}\omega_m^2\tau^2 \nonumber\\
    &&h^{\rm CDI}=\frac{\Omega_c}{\Omega_m}\omega_m\tau-\frac{3}{8}\frac{\Omega_c}{\Omega_m}\omega_m^2\tau^2\nonumber\\
     &&\delta_\gamma^{\rm CDI}=\delta_\nu^{\rm CDI}=-\frac{2}{3}\frac{\Omega_c}{\Omega_m}\omega_m\tau+\frac{1}{4}\frac{\Omega_c}{\Omega_m}\omega_m^2\tau^2\nonumber\\
    &&\delta_c^{\rm CDI}-1=\delta_b^{\rm CDI}=\frac{3}{4}\delta_\gamma^{\rm CDI}\nonumber\\
    &&\theta_c^{\rm CDI}=0\nonumber\\
    &&\theta_\gamma^{\rm CDI}=\theta_b^{\rm CDI}=\theta_{\nu}^{\rm CDI}=-\frac{1}{12} \frac{\Omega_c}{\Omega_m}\omega_m k^2\tau^2\nonumber\\
    &&\sigma_{\nu}^{\rm CDI}=\frac{-1}{6(15+2R_{\nu})}\frac{\Omega_c}{\Omega_m}\omega_m k^2\tau^3,\\
    &&\eta^{\rm BDI}= -\frac{1}{6}\frac{\Omega_b}{\Omega_m}\omega_m\tau+\frac{1}{16}\frac{\Omega_b}{\Omega_m}\omega_m^2\tau^2 \nonumber\\
    &&h^{\rm BDI}=\frac{\Omega_b}{\Omega_m}\omega_m\tau-\frac{3}{8}\frac{\Omega_b}{\Omega_m}\omega_m^2\tau^2\nonumber\\
    &&\delta_\gamma^{\rm BDI}=\delta_\nu^{\rm BDI}=-\frac{2}{3}\frac{\Omega_b}{\Omega_m}\omega_m\tau+\frac{1}{4}\frac{\Omega_b}{\Omega_m}\omega_m^2\tau^2\nonumber\\
     &&\delta_b^{\rm BDI}-1=\delta_c^{\rm BDI}=\frac{3}{4}\delta_\gamma^{\rm BDI}\nonumber\\
    &&\theta_c^{\rm BDI}=0\nonumber\\
    &&\theta_\gamma^{\rm BDI}=\theta_b^{\rm BDI}=\theta_{\nu}^{\rm BDI}=-\frac{1}{12} \frac{\Omega_b}{\Omega_m}\omega_m k^2\tau^2\nonumber\\
    &&\sigma_{\nu}^{\rm BDI}=\frac{-1}{6(15+2R_{\nu})}\frac{\Omega_b}{\Omega_m}\omega_m k^2\tau^3,
\end{eqnarray}
where $\omega_{m}\equiv \sqrt{8\pi G/3}\, a(\tau_{\rm ini})\bar\rho_m(\tau_{\rm ini})/\sqrt{\bar\rho_{r}(\tau_{\rm ini})}$ and $\bar\rho_m$,$\bar\rho_r$ is the background matter and radiation density respectively. $\Omega_b$ ($\Omega_m$) is the factional energy density in baryons (total matter) today. At the time initial conditions are set, $\omega_m\tau\ll 1$ can be treated as an expansion parameter.

\subsection{NDI}
The set of the NDI mode is given by
\begin{eqnarray}
    &&\eta^{\rm NDI}= \frac{-R_{\nu}}{6(15+4R_{\nu})}(k\tau)^2 \nonumber\\
    &&h^{\rm NDI}=\frac{R_{\nu}}{40(1-R_{\nu})}\frac{\Omega_b}{\Omega_m}\omega_m k^2\tau^3\nonumber\\
    &&\delta_\gamma^{\rm NDI}=\frac{-R_{\nu}}{1-R_{\nu}}\left(1-\frac{1}{6}(k\tau)^2\right)\nonumber\\
     &&\delta_b^{\rm NDI}=\frac{R_{\nu}}{8(1-R_{\nu})}(k\tau)^2\nonumber\\
      &&\delta_c^{\rm NDI}=\frac{-R_{\nu}}{80(1-R_{\nu})}\frac{\Omega_b}{\Omega_m}\omega_m  k^2\tau^3\nonumber\\
&&\delta_{\nu}^{\rm NDI}=1-\frac{1}{6}(k\tau)^2\nonumber\\
    &&\theta_\gamma^{\rm NDI}=\theta_b^{\rm NDI}=\frac{-R_{\nu}}{4(1-R_{\nu})}k^2\tau+\frac{3\Omega_b}{16\Omega_m(1-R_{\nu})^2}\omega_m k^2\tau^2\nonumber\\
    &&\theta_c^{\rm NDI}=0\nonumber\\
    &&\theta_{\nu}^{\rm NDI}=\frac{1}{4}k^2\tau\nonumber\\
    &&\sigma_{\nu}^{\rm NDI}=\frac{1}{2(15+4R_{\nu})}k^2\tau^2,
\end{eqnarray}
where $R_\nu\equiv \bar \rho_\nu/\bar\rho_{r}$.
\subsection{DRDI}
We assume DR is free-streaming, analogous to Standard Model neutrinos. We keep terms up to ${\cal O}((k\tau)^2)$. For terms that are zero up to this order, we retain the leading non-vanishing term. The initial conditions are
\begin{eqnarray}
    &&\eta^{\rm DRDI}= \frac{-R_{\rm dr}+R_{\rm dr}^2+R_{\rm dr}R_{\nu}}{6(1-R_{\rm dr})(15+4R_{\rm dr}+4R_{\nu})}(k\tau)^2 \nonumber\\
    &&h^{\rm DRDI}=\frac{R_{\rm dr}}{40(1-R_{\rm dr})}\frac{\Omega_b}{\Omega_m}\omega_m k^2\tau^3\nonumber\\
    &&\delta_\gamma^{\rm DRDI}=\delta_\nu^{\rm DRDI}=\frac{-R_{\rm dr}}{1-R_{\rm dr}}\left(1-\frac{1}{6}(k\tau)^2\right)\nonumber\\
     &&\delta_b^{\rm DRDI}=\frac{R_{\rm dr}}{8(1-R_{\rm dr})}(k\tau)^2\nonumber\\
      &&\delta_c^{\rm DRDI}=\frac{-R_{\rm dr}}{80(1-R_{\rm dr})}\frac{\Omega_b}{\Omega_m}\omega_m  k^2\tau^3\nonumber\\
&&\delta_{\rm dr}^{\rm DRDI}=1-\frac{1}{6}(k\tau)^2\nonumber\\
    &&\theta_\gamma^{\rm DRDI}=\theta_b^{\rm DRDI}=\frac{-R_{\rm dr}}{4(1-R_{\rm dr})}k^2\tau\nonumber\\
    &&\theta_{c}^{\rm DRDI}=0\\
    &&\theta_{\rm dr}^{\rm DRDI}=\frac{1}{4}k^2\tau\nonumber\\
    &&\sigma_{\nu}^{\rm DRDI}=\frac{-19R_{\rm dr}}{30(1-R_{\rm dr})(15+4R_{\rm dr}+4R_{\nu})}k^2\tau^2\nonumber\\
     &&\sigma_{\rm dr}^{\rm DRDI}=\frac{15-15R_{\rm dr}+4R_{\nu}}{30(1-R_{\rm dr})(15+4R_{\rm dr}+4R_{\nu})}k^2\tau^2,\nonumber
\end{eqnarray}
where $R_{\rm dr}\equiv\bar\rho_{\rm dr}/\bar\rho_{r}$.

\bibliography{references}
\end{document}